\documentclass[runningheads]{llncs}
\usepackage{graphicx}
\usepackage{times}
\usepackage{url}
\usepackage{latexsym}
\usepackage{times}
\usepackage{todonotes}
\usepackage{listings}
\usepackage{hyperref}
\usepackage{subfig}
\usepackage{multirow}
\usepackage{comment}
\usepackage{url}
\usepackage{subfig}
\usepackage{amssymb}
\usepackage{hhline}
\usepackage{tabularx}
\usepackage{enumitem}

\begin{document}
%
\title{Coreference Resolution in Research Papers from Multiple Domains}
%
%
\author{Arthur Brack\inst{1}\orcidID{0000-0002-1428-5348}
\and Daniel Uwe M\"uller\inst{1}\orcidID{0000-0002-4492-5879}
\and \\ Anett Hoppe\inst{1}\orcidID{0000-0002-1452-9509}
\and Ralph Ewerth\inst{1,2}\orcidID{0000-0003-0918-6297}}

\authorrunning{A. Brack et al.}
%
\institute{TIB -- Leibniz Information Centre for Science and Technology, Hannover, Germany \and
L3S Research Center, Leibniz University, Hannover, Germany \\
\email{\{arthur.brack|anett.hoppe|ralph.ewerth\}@tib.eu}}

%
%
%
\maketitle              
\begin{abstract}
Coreference resolution is essential for automatic text understanding to facilitate high-level information retrieval tasks such as text summarisation or question answering. Previous work indicates that the performance of state-of-the-art approaches (e.g. based on BERT) noticeably declines when applied to scientific papers. In this paper, we investigate the task of coreference resolution in research papers and subsequent knowledge graph population. We present the following contributions: (1)~We annotate a corpus for coreference resolution that comprises 10 different scientific disciplines from Science, Technology, and Medicine (STM); (2)~We propose transfer learning for automatic coreference resolution in research papers; (3)~We analyse the impact of coreference resolution on knowledge graph (KG) population; (4)~We release a research KG that is automatically populated from 55,485 papers in 10 STM domains. Comprehensive experiments show the usefulness of the proposed approach. Our transfer learning approach considerably outperforms state-of-the-art baselines on our corpus with an F1 score of 61.4 (+11.0), while the evaluation against a gold standard KG shows that coreference resolution improves the quality of the populated KG significantly with an F1 score of 63.5 (+21.8).

\end{abstract}

\section{Introduction}
\label{intro}

Current research is generally published in form of PDF files and, sometimes, research artefacts of other modalities (data sets, source code, etc.). This makes them hard to handle for retrieval systems, since their content is hidden in human- but not machine-interpretable text. In consequence, current academic search engines are not able to adequately support researchers in their day-to-day tasks. This is further aggravated by the exploding number of published articles \cite{bornmann15growth}. 

Approaches to automatically structure research papers are thus an active area of research. \emph{Coreference resolution} is the task to identify mentions in a text which refer to the same entity or concept. It is an essential step for automatic text understanding and facilitates down-stream tasks such as text summarisation or question answering. For instance, the text \textit{`Coreference resolution is... It is used for question answering...'}, has two coreferent mentions \textit{`Coreference resolution'} and \textit{`It'}. This allows us to extract the fact \textless coreference resolution, used\_for, question answering\textgreater.

Current methods for coreference resolution based on deep learning achieve quite impressive results (e.g. an F1 score of 79.6 for the OntoNotes 5.0 dataset~\cite{Joshi2020SpanBERT}) in the general domain, that is data from phone conversations, news, magazines, etc. 
But results of previous work indicate \cite{CohenLCBBPVPH17Craft,KimNWTTY12,NguyenKMMT12,SchaferSS12CorefACL} that general coreference resolution systems perform poorly on scientific text. This is presumably caused by the specific terminology and phrasing used in a scientific domain. 
Some other studies state that annotating scientific text is costly since it demands certain expertise in the article’s domain~\cite{augenstein2017semeval,Brack2020DomainindependentEO,gabor2018semeval}.
Most corpora for research papers cover only a single domain (e.g. bio\-medicine~\cite{CohenLCBBPVPH17Craft}, 
artificial intelligence~\cite{Luan2018MultiTaskIO}) and are thus limited to these domains. 
As a result, the annotated corpora are relatively small and overall only a few domains are covered. 
Datasets for the general domain are usually much larger, but they have not been exploited yet by approaches for coreference resolution in research papers.

Coreference resolution is also one of the main steps in the KG population pipeline \cite{LubaniNM19,Pujara2018MiningKG}. 
However, to date it is not clear, to which extent (a)~coreference resolution can help to reduce the number of scientific concepts in the populated KG, and (b)~how coreference resolution influences the quality of the populated KG. 
Besides, a KG comprising multiple scientific domains has not been populated yet.

In this paper, we address the task of coreference resolution in research papers and subsequent knowledge graph population. 
Our contributions can be summarised as follows:
(1) First, we annotate a corpus for coreference resolution that consists of 110 abstracts from 10 domains from Science, Technology, and Medicine. The systematic annotation resulted in a substantial inter-coder agreement (0.68 $\kappa$). 
We provide and compare baseline results for this dataset by evaluating five different state-of-the-art approaches: (i) coreference resolution systems for the general~\cite{Joshi2020BFCR} and (ii) for the artificial intelligence domain~\cite{Luan2018MultiTaskIO};
(iii) supervised learning with training data from our corpus with a SpanBERT~\cite{Joshi2020SpanBERT} and (iv)~a SciBERT-based system~\cite{Beltagy2019SciBERTPC}, and (v) the Scientific Information Extractor~\cite{Luan2018MultiTaskIO}.
Our experimental results confirm that state-of-the-art coreference approaches do not perform well on research papers.
(2) Consequently, we propose sequential transfer learning for coreference resolution in research papers. This approach utilises our corpus by fine-tuning a model that is pre-trained on a large corpus from the general domain~\cite{Pradhan2013TowardsRL}.
Experimental results show that our approach significantly outperforms the best state-of-the-art baseline (F1 score of 61.4, i.e. +11.0). 
(3) We investigate the impact of coreference resolution on automatic KG population. To evaluate the quality of various KG population strategies, we (i) compile a gold standard KG from our annotated corpus that contains scientific concepts referenced by mentions from text, and (ii)~present a procedure to evaluate the clustering results of mentions.
(4) We release (i) an automatically populated KG from 55,485 abstracts of the 10 STM domains and (ii) a gold KG (Test-STM-KG) from the annotated STM-corpus. 
Experimental results show that coreference resolution has only a small impact on the number of concepts in a populated KG, but it helps to improve the quality of the KG significantly: the population with coreference resolution yields an F1 score of 63.5 
evaluated against the gold KG (+21.8 F1).
We release all our corpora and source code to facilitate further research.

The remainder of the paper is organised as follows: Section 2 summarises related work on coreference resolution.
Section~3 describes the annotation procedure and the characteristics of the corpus, and our proposed approaches for coreference resolution, KG population and KG evaluation.
The experimental setup and results are reported in Section 4 and 5, while Section 6 concludes the paper and outlines future work.

\section{Related Work}
\label{sec:related_work}

\subsection{Approaches for Coreference Resolution}
For a given document $d$, the task of coreference resolution is (a) to extract mentions of scientific concepts $M(d) = \{m_1, ..., m_h\}$, and (b) to cluster mentions that refer to the same concept, i.e. $c_d(m) \subseteq M(d)$ is the cluster for mention $m$. 
Recent approaches mostly rely on supervised learning and can be categorised into three groups~\cite{Ng2017Coreference}:
(1)~Men\-tion-pair models~\cite{NgC02Coreference,SoonNL01Coref} are binary classifiers that determine whether two mentions are coreferent or not. 
(2)~Entity-mention models~\cite{ClarkM15Coref,RahmanN09Coref} determine whether a mention is coreferent to a preceding \emph{cluster}. A cluster has more expressive features compared to a mention in mention-pair models. 
(3)~Ranking-based models~\cite{DenisB08Coref,Lee2017EndtoendNC,MarasovicBOF17Coref} simultaneously rank all candidate antecedents (i.e. preceding mention candidates). This enables the model to identify the most probable antecedent.  

Lee et al.~\cite{Lee2017EndtoendNC,LeeHZ18Coreference} propose an end-to-end neural coreference resolution model. It is a ranking-based model that jointly recognises mentions and clusters. Therefore, the model considers all spans in the text as possible mentions and learns distributions over possible antecedents for each mention. For computational efficiency, candidate spans and antecedents are pruned during training and inference.
Joshi et al.~\cite{Joshi2020BFCR} enhance Lee et al.'s model with BERT-based word embeddings~\cite{Devlin2018BERTPO}, while Ma et al.~\cite{Ma0LHPSL20JointlyCoreference} improve the model with better attention mechanisms and loss functions. 

Furthermore, several approaches proposed multi-task learning, such that related tasks may benefit from knowledge in other tasks to achieve better prediction accuracy:
Luan et al.~\cite{Luan2018MultiTaskIO,WaddenWLH19DyGIE} train a model on three tasks (coreference resolution, entity and relation extraction) using one dataset of research papers. 
Sanh et al.~\cite{Sanh2018AHM} introduce a multi-task model that is trained on four tasks (mention detection, coreference resolution, entity and relation extraction) using two different datasets in the general domain. 

Results of some previous studies~\cite{CohenLCBBPVPH17Craft,NguyenKMMT12,KimNWTTY12,SchaferSS12CorefACL} revealed that general coreference systems do not work well in the biomedical domain due to the lack of domain knowledge. For instance, on Colorado Richly Annotated Full Text (CRAFT) corpus~\cite{CohenLCBBPVPH17Craft} a coreference resolution system for the news domain achieves only 14.0 F1 (-32.0).

To the best of our knowledge, a transfer learning approach from the general to the scientific domain has not been proposed for coreference resolution yet.

\subsection{Corpora for Coreference Resolution in Research Papers}
For the general domain, multiple datasets exist for coreference resolution, e.g. Message Understanding Conference (MUC-7)~\cite{Mikheev1998SeventhMU}, Automatic Content Extraction (ACE05)~\cite{Doddington2004TheAC}, or OntoNotes 5.0~\cite{Pradhan2013TowardsRL}. The OntoNotes 5.0 dataset~\cite{Pradhan2013TowardsRL} is the largest one and is used in many benchmark experiments for coreference resolution systems~\cite{Lee2017EndtoendNC,Joshi2020BFCR,Ma0LHPSL20JointlyCoreference}. 

Various annotated datasets for coreference resolution exist also for research papers:
CRAFT corpus~\cite{CohenLCBBPVPH17Craft} covers 97 papers from biomedicine.
The corpus of Schäfer et al.~\cite{SchaferSS12CorefACL} contains 266 papers from computational linguistics and language technology.
Chaimongkol et al. \cite{ChaimongkolAT14Corpus} annotated a corpus of 284 papers from four subdisciplines in computer science.
The SciERC corpus~\cite{Luan2018MultiTaskIO} comprises 500 abstracts from the artificial intelligence domain and features annotations for scientific concepts and relations. It was used to generate an artificial intelligence (AI) knowledge graph~\cite{Dessi2020AIKG}.
Furthermore, several datasets exist for scientific concept extraction~\cite{augenstein2017semeval,Luan2018MultiTaskIO,Brack2020DomainindependentEO,handschuh2014acl} and relation extraction~\cite{Luan2018MultiTaskIO,gabor2018semeval,augenstein2017semeval} that cover various scientific domains.

To the best of our knowledge, a corpus for coreference resolution that comprises a broad range of scientific domains is not available yet.

\section{Coreference Resolution in Research Papers}

As the discussion of related work reveals, existing corpora for coreference resolution in scientific papers normally cover only a single domain, and coreference resolution approaches do not perform well on scholarly texts. To address these issues, we systematically annotate a corpus with coreferences in abstracts from 10 different science domains.
Current approaches for coreference resolution in research papers do not exploit existing annotated datasets from the general domain, which are usually much larger than in the scientific domain. We propose a sequential transfer learning approach that takes advantage from large, annotated datasets.
Finally, to the best of our knowledge, the impact of (a) coreference resolution and (b) cross-domain collapsing of mentions to scientific concepts on KG population with multiple science domains has not been investigated yet. Consequently, we present an evaluation procedure for the clustering aspect in the KG population pipeline.

In the sequel, we describe our annotated corpus, our transfer learning approach for coreference resolution, and an evaluation procedure for clustering in KG population.

\subsection{Corpus for Coreference Resolution in 10 STM Domains}
In this section, we describe the STM corpus~\cite{Brack2020DomainindependentEO}, which we used as the basis for the annotation, our annotation process, and the characteristics of the resulting corpus.

\paragraph{STM Corpus:}
The STM corpus~\cite{Brack2020DomainindependentEO} comprises 110  articles from 10 domains in Science, Technology and Medicine, namely Agriculture (Agr), Astronomy (Ast), Biology (Bio), Chemistry (Che), Computer Science (CS), Earth Science (ES), Engineering (Eng), Materials Science (MS), Mathematics (Mat), and Medicine (Med). 
It contains annotated mentions of scientific concepts in abstracts with four domain-independent concept types, namely \textit{Process}, \textit{Method}, \textit{Material}, and \textit{Data}. These concept mentions were later linked to entities in Wikipedia and Wikidata~\cite{DSouza2020STEM}. 
The 110 articles (11 per domain) were taken from the OA-STM corpus~\cite{OASTM} of Elsevier Labs.

We build upon related work and extend the STM corpus with coreference annotations. In particular, we (1) annotate coreference links between existing scientific concept mentions in abstracts using the BRAT annotation tool~\cite{StenetorpPTOAT12BRAT}, and (2) annotate further mentions, i.e. pronouns and noun phrases consisting of multiple consecutive mentions.

\paragraph{Annotation Process:}
Other studies have shown that non-expert annotations are viable for the scientific domain~\cite{Brack2020DomainindependentEO,Chambers2013,Fisas2015OnTD,SchaferSS12CorefACL,teufel2009towards}, 
and they are less costly than domain-expert annotations.
Therefore, we also annotate the corpus with non-domain experts, i.e. by two students in computer science.
Furthermore, we follow mostly the annotation procedure of the STM corpus~\cite{Brack2020DomainindependentEO}, which consists of the following three phases:
\begin{enumerate}[nosep]

\item \emph{Pre-Annotation:}
This phase aims at developing annotation guidelines through trial annotations.
We adapted the comprehensive annotation guidelines of the OntoNotes 5.0 dataset~\cite{PradhanCoNLL2012}, which were developed for the general domain, to research papers. 
In particular, we provide briefer and simpler descriptions with examples from the scientific domain.
Within three iterations both annotators labelled independently 10, 9 and 7 abstracts (i.e. 26 abstracts), respectively.
After each iteration the annotators discussed the outcome and refined the annotation guidelines.

\item \emph{Independent Annotation:}
After the annotation guidelines were finalised, both annotators independently re-annotated  the previously annotated abstracts and 24 additional abstracts. 
The final inter-coder agreement was measured on the 50 abstracts (5 per domain) using Cohen's $\kappa$ \cite{cohen1960coefficient,Kopec2014} and MUC~\cite{VilainBACH95MUC}.
As shown in Table~\ref{table:ia_agreement}, we achieve a substantial agreement with 0.68 $\kappa$ and 0.69 MUC. 

\item \emph{Consolidation:}
Finally, the remaining 60 abstracts were annotated by one annotator and the annotation results of this author were used as the gold standard corpus. 

\end{enumerate}

\begin{table}[tb]
\centering
\small
\caption{Per-domain and overall inter-annotator agreement (Cohen's $\kappa$ and MUC) for coreference resolution annotation in our STM corpus.}
\begin{tabular}{l|rrrrrrrrrr|r}
	& \textit{Mat}	&\textit{Med}	&\textit{Ast}	&\textit{CS}	&\textit{Bio}	&\textit{Agr}	&\textit{ES}	&\textit{Eng}	&\textit{Che}	&\textit{MS} &\textit{Overall} \\ \hline
$\kappa$	& 0.84 & 0.80 & 0.78 & 0.72 & 0.70 & 0.66 & 0.61 & 0.58 & 0.56 & 0.52 & 0.68 \\ 
MUC & 0.83 & 0.69 & 0.78 & 0.73 & 0.70 & 0.72 & 0.61 & 0.66 & 0.56 & 0.63 & 0.69 \\
\end{tabular}
\label{table:ia_agreement}
\vspace{-1em}
\end{table}

\begin{table}[tb]
\centering
\small
\caption{Characteristics of the annotated STM corpus with 110 abstracts per concept type in terms of number of scientific concept mentions, number of coreferent mentions, number of coreference clusters and singleton clusters, and the number of overall clusters. MIXED denotes clusters consisting of mentions with different concept types, NONE denotes coreference mentions and clusters without a scientific concept mention.}
\label{tab:characteristics_per_concept}
\begin{tabular}{l|rrrr|rr|r}
                      & Data  & Material & Method & Process & MIXED  & NONE  & Total  \\ \hline
\# mentions            & 1,658 & 2,099 & 258 & 2,112 & 0  & 0   & 6,127 \\
\# coreferent mentions & 351  & 910  & 101 & 510  & 0  & 705 & 2,577 \\ \hline
\# coreference clusters            & 153  & 339  & 30  & 198  & 50 & 138 & 908  \\
\# singleton clusters          & 1,307 & 1,189 & 157 & 1,602 & 0  & 0   & 4,255 \\ \hline
\# overall clusters            & 1,460 & 1,528 & 187 & 1,800 & 50 & 138 & 5,163 \\ 
\end{tabular}
\end{table}

\begin{table}[t!]
\centering
\small
\caption{Characteristics of the STM corpus per domain (11 abstracts per domain).}
\label{tab:corpus_characteristics_per_domain}
\begin{tabular}{l|rrrrrrrrrr|r}
                  & Agr & Ast & Bio & Che & CS  & ES  & Eng & MS  & Mat & Med & Total \\ \hline
\# mentions           & 741 & 791 & 649 & 553 & 483 & 698 & 741 & 574 & 297 & 600 & 6,127 \\
\# coreferent mentions & 276 & 365 & 275 & 282 & 181 & 241 & 318 & 256 & 124 & 259 & 2,577 \\ \hline
\# coreference clusters            & 106 & 120 & 98  & 90  & 67  & 93  & 117 & 87  & 48  & 82  & 908  \\
\# singleton clusters          & 520 & 549 & 443 & 384 & 339 & 525 & 503 & 371 & 210 & 411 & 4,255 \\ \hline
\# clusters            & 626 & 669 & 541 & 474 & 406 & 618 & 620 & 458 & 258 & 493 & 5,163 \\ 
\end{tabular}
\vspace{-1em}
\end{table}

\paragraph{Corpus Characterstics:}

Table~\ref{tab:characteristics_per_concept} shows the characteristics of the resulting corpus broken down per concept type, while they are listed per domain in Table~\ref{tab:corpus_characteristics_per_domain}. The original corpus has in total 6,127 mentions. 2,577 mentions were annotated as coreferent resulting in 908 coreference clusters. 
Thus, each coreference cluster contains on average 2.84 mentions, while \emph{Method} clusters contain the most (3.4 mentions) and \emph{Data} clusters the least (2.3 mentions).
Furthermore,  705 mentions were annotated additionally (referred to as NONE) since they represent pronouns (422 mentions) or noun phrases consisting of multiple consecutive original mentions (283 mentions) such as \textit{`... [[A], [B], and [C] [treatments]]... [These treatments]...'}.
Fifty clusters (5\%) contain mentions with different concept types (referred to as MIXED) due to disagreements between the annotators of the original concept mentions, and the annotators of coreferences. For instance, non-coreferent mentions were annotated as coreferent, or coreferent mentions have different concept types.
Finally, 138 clusters (15\%) do not have a concept type (NONE) since they form clusters which are not coreferent with the original concept mentions.

\subsection{Transfer Learning for Coreference Resolution}

We suggest sequential transfer learning~\cite{Ruder2019Neural} for coreference resolution in research papers. Therefore, we fine-tune a model pre-trained on a large (source) dataset to our (target) dataset. 
As the source dataset, we use the English portion of the OntoNotes 5.0 dataset~\cite{Pradhan2013TowardsRL}, since it is a broad corpus that consists of 3,493 documents with telephone conversations, magazine and news articles, web data, broadcast conversations, and the New Testament. Besides, our annotation guidelines were adapted from OntoNotes~5.0.

For the model, we utilise \emph{BERT for Coreference Resolution (BFCR)}~\cite{Joshi2020BFCR} with \emph{SpanBERT}~\cite{Joshi2020SpanBERT} word embeddings. This model achieves state-of-the-art results on the Onto\-Notes dataset~\cite{Joshi2020SpanBERT}. Another advantage is the availability of the pre-trained model and the source code.
The BFCR model improves Lee et al.'s approach~\cite{LeeHZ18Coreference} by replacing the LSTM encoder with the SpanBERT transformer-encoder. SpanBERT~\cite{Joshi2020SpanBERT} has different training objectives than BERT~\cite{Devlin2018BERTPO} to better represent spans of text.

\subsection{Cross-Domain Research Knowledge Graph Population}
\label{sec:kg_construction_approach}

Let $d \in D$ be an abstract, $M(d) = \{m_1, ..., m_h\}$ the mentions of scientific concepts in $d$, and $c_d(m_i) \subseteq M(d)$ the corresponding coreference cluster for mention $m_i$ in $d$. If mention $m_s$ is not coreferent with other mentions in $d$, then $c_d(m_s) = \{m_s\}$ is a singleton cluster. The set of all clusters is denoted by $C$.
An equivalence relation $collapsable \subseteq C\times C$ defines if two clusters can be collapsed, i.e. if the clusters refer to the same scientific concept. 
To create the set of all concepts $E$, we build the quotient set for the set of clusters $C$ with respect to the relation $collapsable$:
\begin{eqnarray}
C := \{c_d(m) | d \in D, m \in M(d)\} \\
{[c]} := \{x \in C | collapsable(c, x)\} \\
E := \{[c] | c \in C\} 
\end{eqnarray}
Now, we can construct the KG:
for each paper $d \in D$ and for each scientific concept $e \in E$ we create a node in the KG.
The scientific concept type of $e$ is the most frequent concept type of all mentions in $e$.
Then, for each mention $m \in M(d)$ we create a `mentions' link between the paper and the corresponding scientific concept $[m] \in E$.

\paragraph{Cross-Domain vs. In-Domain Collapsing:}
One commonly used approach to define the $collapsable$ relation is to treat two clusters as equivalent, if and only if the `label' of the clusters is the same.
The label of a cluster is the longest mention in the cluster normalised by (a) lower-casing, (b) removing articles, possessives and demonstratives, (c) resolving acronyms, and (d) lemmatisation using WordNet~\cite{Fellbaum2000WordNetA} to transform plural forms to singular. 
Other studies \cite{Dessi2020AIKG,Luan2018MultiTaskIO} used a similar label function for KG population.

However, a research KG that comprises multiple scientific disciplines has not been populated yet. 
Thus, it is not clear whether it is feasible to collapse clusters across domains. Usually, terms within a scientific domain are unambiguous. However, some terms have different meanings across scientific disciplines (e.g. ``neural network" in \emph{CS} and \emph{Med}).
Thus, we investigate both cross-domain and in-domain collapsing strategies.

\paragraph{Knowledge Graph Population Approach:}
We populate a research KG with research papers from multiple scientific domains, i.e. 55,485 abstracts of Elsevier with CC-BY licence from the 10 investigated domains.
First, we extract (a) concept mentions from the abstracts using the scientific concept extractor of the STM-corpus~\cite{Brack2020DomainindependentEO}, and (b) clusters within the abstracts with our transfer learning coreference model.
Then, those mention clusters, which contain solely mentions recognised by the coreference resolution model and not by the scientific concept extraction model, are dropped, since the coreference resolution model does not recognise the concept type of the mentions.
Finally, the remaining clusters serve for the population of the KG as described above.

\subsection{Evaluation Procedure of Clustering in KG Population}

One common approach to evaluate the quality of a populated KG is to annotate a (random) subset of statements by humans as true or false and to calculate precision and recall~\cite{Dessi2020AIKG,Weikum2020MachineKnowledge}. To evaluate recall, small collections of ground-truth capturing \emph{all} knowledge is necessary, that are usually difficult to obtain~\cite{Weikum2020MachineKnowledge}.
To the best of our knowledge, a common approach to evaluate the clustering aspect of the KG population pipeline does not exist yet.
Thus, in the following, we present (1) an annotated test KG, and (2) metrics to evaluate clustering of mentions to concepts in KG population.

\paragraph{Test KG:}
To enable evaluation of KG population strategies, we compile a test KG, referred to as \emph{Test-STM-KG}.
For this purpose, we reuse the STEM-ECR corpus~\cite{DSouza2020STEM}, in which 1,221 mentions of the STM corpus are linked to Wikipedia entities.
First, we extract all annotated clusters of the STM corpus in which all mentions of the cluster uniquely refer to the same Wikipedia entity. 
Then, we collapse all clusters which refer to the same Wikipedia entity to concepts.
Formally, the Test-STM-KG is a partition of mentions, where each part denotes a concept, i.e. a disjoint set of mentions. 
A mention is uniquely represented by the tuple (start offset, end offset, concept type, doc id).

Table~\ref{tab:test-stm-kg} shows the characteristics of the compiled Test-STM-KG.
It consists of 920 clusters, of which 711 are singleton clusters. These clusters were collapsed to 762 concepts,
of which 31 concepts are used across multiple domains (referred to as MIX).

\begin{table}[tb]
\centering
\small
\caption{Characteristics of the \emph{Test-STM-KG}: number of concepts per concept type and per domain. MIX denotes the number of cross-domain concepts.}
\label{tab:test-stm-kg}
\begin{tabular}{l|rrrrrrrrrrr|r}
         & Agr & Ast & Bio & CS & Che & ES & Eng & MS & Mat & Med & MIX & Total \\ \hline
Data     & 5   & 18  & 3   & 20 & 4   & 9  & 28  & 13 & 37  & 8   & 9  & 154   \\
Material & 27  & 35  & 30  & 20 & 26  & 52 & 32  & 30 & 9   & 40  & 7  & 308   \\
Method   & 1   & 1   & 1   & 21 & 6   & 2  & 4   & 10 & 3   & 8   & 7  & 64    \\
Process  & 17  & 12  & 21  & 34 & 13  & 33 & 20  & 25 & 15  & 38  & 8  & 236   \\ \hline
Total    & 50  & 66  & 55  & 95 & 49  & 96 & 84  & 78 & 64  & 94  & 31 & 762  
\end{tabular}
\end{table}

\paragraph{Evaluation Procedure:}
\label{sec:eval_procedure_kg_population}

To evaluate the clustering result of a KG population strategy, we use the metrics of coreference resolution.
The three popular metrics for coreference resolution are $MUC$ \cite{VilainBACH95MUC}, $B^3$ \cite{Bagga98B3} and $CEAFe_{\phi 4}$ \cite{Luo05CEAF}. Each of them represents different evaluation aspects (see \cite{Pradhan2014Scoring} for more details). 
To calculate these metrics, we treat the gold concepts (i.e. a partition of mentions) of the Test-STM-KG as the `key' and the predicted concepts as the `response'.
We report also the \emph{CoNLL~P/R/F1} scores, that is the averages of $MUC$'s, $B^3$'s and $CEAFe_{\phi 4}$'s respective precision~(P), recall~(R) and F1 scores. The CoNLL metrics were proposed for the conference on Computational Natural Language Learning (CoNLL) shared tasks on coreference resolution~\cite{Pradhan2014Scoring}.

\section{Experimental Setup}
Here we describe our experimental setup for coreference resolution and KG population.

\subsection{Automatic Coreference Resolution}

We evaluate three different state-of-the-art architectures on the STM dataset:
(I) \emph{BERT for Coreference Resolution (BFCR)}~\cite{Joshi2020BFCR} with \emph{SpanBERT}~\cite{Joshi2020SpanBERT} word embeddings (referred to as \emph{BFCR\_Span}), (II) BFCR with \emph{SciBERT}~\cite{Beltagy2019SciBERTPC} word embeddings (referred to as \emph{BFCR\_Sci}), and (III) \emph{Scientific Information Extractor (SCIIE)}~\cite{Luan2018MultiTaskIO} with ELMo~\cite{Peters2018ELMo} word embeddings (referred to as \emph{SCIIE}).
The three architectures are evaluated in the following six approaches (\#1 - \#6):
\begin{itemize}[nosep] 
    \item \emph{Pre-Trained Models:}
We evaluate already pre-trained models on the test sets of the STM corpus, i.e.
\#1 \emph{BFCR\_Span} trained on the English portion of the OntoNotes dataset~\cite{PradhanCoNLL2012}, and \#2 \emph{SCIIE} trained on SciERC~\cite{Luan2018MultiTaskIO} from the AI domain.    

    \item \emph{Supervised Learning:} 
We train a model from scratch with the three architectures using the training data of the STM corpus and evaluate their performance with the test sets of STM:
\#3 \emph{BFCR\_Span}, \#4 \emph{BFCR\_Sci}, and \#5  \emph{SCIIE}.    

    \item \emph{Transfer Learning:} 
This is our proposed approach \#6.
We fine-tune all parameters of a pre-trained model on the English portion of the OntoNotes dataset~\cite{Joshi2020SpanBERT} with the training data of our STM corpus.
For that, we use the \emph{BFCR\_Span} architecture.    
\end{itemize}

\paragraph{Evaluation:}
We use the metrics $MUC$ \cite{VilainBACH95MUC}, $B^3$ \cite{Bagga98B3}, $CEAFe_{\phi 4}$ \cite{Luo05CEAF} and $CoNLL$ \cite{Pradhan2014Scoring} in compliance with other studies on coreference resolution~\cite{Joshi2020BFCR,Ma0LHPSL20JointlyCoreference,Lee2017EndtoendNC}.
To obtain robust results, we apply five-fold cross-validation, according to the data splits given by Brack et al. \cite{Brack2020DomainindependentEO}, and report averaged results.
For each fold, the dataset is split into train/validation/test sets with 8/1/2 abstracts per domain, respectively, i.e. 80/10/20 abstracts.  
We reuse the original implementations and default hyperparameters of the above architectures. 
Hyperparameter-tuning of the best baseline approach \#3 according to \cite{Joshi2020BFCR} confirmed that the default hyperparameters of \emph{BFCR\_Span} perform best on our corpus.

\subsection{Evaluation of KG Population Strategies}

We compare four KG population strategies: (1) cross-domain and (2) in-domain collapsing, as well as (3) cross-domain and (4) in-domain collapsing without coreference resolution. 
To evaluate cross-domain and in-domain collapsing, we take the gold clusters (i.e. mention clusters within the abstracts) of the Test-STM-KG and collapse them to concepts according to the respective strategy.
When leaving out the coreference resolution step, we treat all mentions in the Test-STM-KG as singleton clusters and collapse them to concepts according to the respective strategy. Finally, we calculate the metrics as described in Section~\ref{sec:eval_procedure_kg_population}.

\section{Results and Discussion}
In this section, we discuss the experimental results for automatic coreference resolution and KG population. 

\subsection{Automatic Coreference Resolution}

\begin{table}[tb]
\centering
\small
\caption{Performance of the baseline approaches \#1 - \#5 and our proposed transfer learning approach \#6 on the test sets of the STM corpus across five-fold cross validation.}
\label{tab:coref_results}
\begin{tabular}{lll|rrr|rrr|rrr|rrr}
  &           &                & \multicolumn{3}{c}{$MUC$} & \multicolumn{3}{c}{$B^3$} & \multicolumn{3}{c}{$CEAFe_{\phi 4}$} & \multicolumn{3}{c}{$CoNLL$} \\
  &           & Training data  & P      & R      & F1     & P      & R     & F1    & P       & R      & F1     & P       & R      & F1     \\ \hline
\#1 & BFCR\_Span & OntoNotes                                                  & 57.1   & 31.1   & 40.2  & 55.9   & 25.7  & 35.2  & 50.2    & 28.1   & 36.0   & 54.4    & 28.3   & 37.1   \\
\#2 & SCIIE     & SciERC                                                     & 13.4   & 4.5    & 6.8   & 13.1   & 4.3   & 6.5   & 18.1    & 6.0    & 9.0    & 14.9    & 4.9    & 7.4    \\ \hline
\#3 & BFCR\_Span & STM                                                        
& 61.6	& 45.6	& 52.3 & 59.8	& 41.5	& 48.8 & 57.9	& 44.4	& 50.0 & 59.8	& 43.8	& 50.4 \\
\#4 & BFCR\_Sci  & STM                                                        
& 61.9	& 40.2	& 48.6 & 59.7	& 36.1	& 44.9 & 61.7	& 36.9	& 46.0 & 61.1	& 37.7	& 46.5 \\
\#5 & SCIIE     & STM                                                        & 60.3   & 45.2   & 51.6  & 57.6   & 41.7  & 48.3  & 56.6    & 43.6   & 49.1   & 58.1    & 43.5   & 49.7   \\ \hline
\textbf{\#6} & \textbf{BFCR\_Span} & \textbf{Onto$\rightarrow$STM}
& \textbf{64.5}	& \textbf{63.5}	& \textbf{63.9} & \textbf{61.0}	& \textbf{60.0}	& \textbf{60.4} & \textbf{60.5}	& \textbf{59.6}	& \textbf{60.0} & \textbf{62.0}	& \textbf{61.0}	& \textbf{61.4} 
\end{tabular}
\vspace{-1em}
\end{table}

\begin{table}[tb]
\centering
\small
\caption{
Per domain and overall CoNLL F1 results of the best baseline \#3 and our transfer learning approach \#6 on the STM corpus across five-fold cross validation.}
\label{tab:results_per_domain}
\begin{tabular}{lll|r|r|r|r|r|r|r|r|r|r|r}
 & &   Training data & Agr  & Ast  & Bio  & Che  & CS   & ES   & Eng  & MS   & Mat  & Med & Overall  \\ \hline
\#3 & BFCR\_Span & STM & 48.0 & 50.5	& 52.2	& 49.0	& 59.1	& 39.6	& 52.8	& 47.6	& 42.5	& 51.0 & 50.4 \\
\textbf{\#6} & \textbf{BFCR\_Span}  & \textbf{Onto$\rightarrow$STM} & \textbf{62.8} & \textbf{61.1} & \textbf{57.5} & \textbf{56.3} & \textbf{74.9} & \textbf{57.5} & \textbf{59.8} & \textbf{52.1} & \textbf{55.7} & \textbf{62.1} & \textbf{61.4}
\end{tabular}
\vspace{-1em}
\end{table}

Table~\ref{tab:coref_results} shows the overall results of the six evaluated approaches and Table~\ref{tab:results_per_domain} the results per domain of the best baseline \#3 and our approach \#6. Our transfer learning approach \#6 \emph{BFCR\_Span} from OntoNotes (Onto)~\cite{Pradhan2013TowardsRL} to STM significantly outperforms the best baseline approach \#3 with an overall CoNLL F1 of 61.4 (+10.0) and a low standard deviation $\pm1.5$ across the five folds.

The approaches \#1 \emph{BFCR\_Span} pre-trained on OntoNotes~\cite{Pradhan2013TowardsRL}, and \#2 \emph{SCIIE} pre-trained on SciERC~\cite{Luan2018MultiTaskIO} achieve a CoNLL F1 score of 37.1 and 7.4, respectively. These scores are quite low compared to the approaches \#3 - \#6 that use training data of the STM corpus. 
This indicates that models pre-trained on existing datasets do not generalise sufficiently well for coreference resolution in research papers.
Models trained only on the STM corpus (i.e. \#3~-~\#5) achieve better results. However, they have quite low recall scores indicating that the size of the training data might not be sufficient to enable the model to generalise well. 
SciBERT \#4, although pre-trained on scientific texts, performs worse than SpanBERT \#3. Presumably the reason is that SpanBERT has approximately 3 times more parameters than SciBERT.
Our transfer learning approach \#6 achieves the best results with quite balanced precision and recall scores.

Furthermore, to evaluate the effectiveness of our transfer learning approach, we compare the best baseline \#3  and our transfer learning approach \#6 also with the SciERC corpus~\cite{Luan2018MultiTaskIO}. The SciERC corpus comprises 500 abstracts from the AI domain.
Since SciERC has around 5 times more training data than STM, we compare the approaches \#3 and \#6 also using only $\frac{1}{5}$th of the training data in SciERC while keeping the original validation and test sets.
It can be seen in Table~\ref{tab:coref_results_scierc} that our transfer learning approach \#6 improves slightly the baseline result using the whole training data with 60.1~F1 (+0.8).
When using only $\frac{1}{5}$th of the training data, our transfer learning approach noticeably outperforms the baseline with 54.2 F1 (+7.1).
Thus, our transfer learning approach can help significantly to improve the performance of coreference resolution in research papers with few labelled data.

\begin{table}[tb]
\centering
\small
\caption{CoNLL scores on the tests sets of the SciERC corpus~\cite{Luan2018MultiTaskIO} across 3 random restarts of the approaches: current state of the art of Luan et al., 
the best baseline approach (\#3), and our transfer learning approach (\#6). We report results using the whole and using only $\frac{1}{5}$th of the training data of SciERC (referred to as $\frac{1}{5}$SciERC).}
\label{tab:coref_results_scierc}
\begin{tabular}{lll|rrr}
  &           & Training data  & P       & R      & F1     \\ \hline
\multicolumn{2}{l}{ Luan et al.~\cite{Luan2018MultiTaskIO}} & SciERC 
 &  52.0 & 44.9 & 48.2 \\ \hline
\#3 & BFCR\_Span & SciERC 
 & 63.3	& 55.7	& 59.3 \\ 
\textbf{\#6} & \textbf{BFCR\_Span} & \textbf{OntoNotes$\rightarrow$SciERC}
& \textbf{63.9}	& \textbf{57.1}	& \textbf{60.1} \\
\hline
\#3 & BFCR\_Span & $\frac{1}{5}$SciERC 
& 63.1	& 39.1	& 47.1 \\
\textbf{\#6} & \textbf{BFCR\_Span} & \textbf{OntoNotes$\rightarrow \frac{1}{5}$SciERC} & \textbf{52.8}	& \textbf{56.7}	& \textbf{54.2} \\
\end{tabular}
\vspace{-1em}
\end{table}

\subsection{Cross-Domain Research KG}
In this subsection, we describe the characteristics of our populated KG and discuss the evaluation results of various KG population strategies.

\subsubsection{Characteristics of the Research KG:}

Table~\ref{tab:kg_concepts_per_domain_type} shows the characteristics of the populated KGs per domain.
The resulting KGs with cross-domain and in-domain collapsing have more than 994,000 and 1.1 Mio. scientific concepts, respectively, obtained from 55,485 abstracts with more than 2,1 Mio. concept mentions and 726,000 coreferent mentions.
\emph{Ast} and \emph{Bio} are the most represented domains, while \emph{CS} and \emph{Mat} are the most underrepresented.

\begin{table}[tb]
\centering
\small
\caption{Characteristics of the populated research KGs per domain: (1) number of abstracts, number of extracted scientific concept mentions and coreferent mentions, (2) the number of scientific concepts for the KG with cross-domain collapsing, (3) in-domain collapsing, (4) cross-domain collapsing but without coreference resolution, and (5) in-domain collapsing but without coreference resolution. Reduction denotes the percentual reduction of mentions to scientific concepts and MIX the cross-domain concepts. }
\label{tab:kg_concepts_per_domain_type}
\resizebox{\textwidth}{!}{
\begin{tabular}{l|rrrrrrrrrrr|r}
         & Agr    & Ast    & Bio    & CS    & Che   & ES    & Eng   & MS    & Mat  & Med    & MIX & Total     \\ \hline
\# abstracts      & 7,731 & 15,053 & 11,109 & 1,216 & 1,234 & 2,352 & 3,049 &  2,258 & 665 &  10,818  & - & 55,485 \\                  
\# mentions            & 332,983 & 370,311 & 423,315 & 45,388 & 46,203 & 129,288 & 127,985 & 86,490 & 20,466 & 586,019  & - & 2,168,448\\
\# coref. men. & 108,579 & 120,942 & 143,292 & 17,674 & 14,059 & 40,974  & 42,654  & 25,820 & 8,510  & 203,884  & - & 726,388\\ \hline
\multicolumn{13}{c}{cross-domain collapsing}\\
KG concepts      & 138,342 & 173,027 & 177,043 & 20,474 & 21,298 & 62,674 & 55,494 & 39,211 & 9,275 & 227,690 & 70,044 & 994,572\\
- Data     & 27,132 & 64,537 & 32,946 & 5,380 & 5,124 & 19,542 & 17,053 & 10,629 & 2,982 & 66,473 & 19,715 & 271,513\\
- Material & 69,534 & 45,296 & 83,627 & 6,242 & 10,154 & 24,322 & 19,689 & 17,276 & 2,406 & 68,141 & 20,812 & 367,499\\
- Method   & 2,992 & 8,819 & 6,135 & 2,001 & 1,055 & 1,776 & 2,953 & 1,605 & 685 & 9,363 & 1,627 & 39,011\\
- Process  & 38,684 & 54,375 & 54,335 & 6,851 & 4,965 & 17,034 & 15,799 & 9,701 & 3,202 & 83,713 & 27,890 & 316,549 \\
reduction & 58\%	& 53\%	& 58\%	& 55\%	& 54\%	& 52\%	& 57\%	& 55\%	& 55\%	& 61\%	&  - & 54\% \\ 
\hline

\multicolumn{13}{c}{in-domain collapsing}\\
KG concepts & 180,135 &	197,605 &	229,201 &	30,736 &	32,191 &	81,584 &	78,417 &	55,358 &	14,567 &	278,686 &- &	1,178,480 \\
reduction & 46\% &	47\% &	46\% &	32\% &	30\% &	37\% &	39\% &	36\% &	29\% &	52\% &	- & 46\% \\ 
\hline

\multicolumn{13}{c}{cross-domain collapsing without coreference resolution}\\
KG concepts	& 146,894 & 	182,479 & 	187,557 & 	21,950 & 	22,555 & 	66,600 & 	59,689 &  	41,776 & 	9,939 & 	242,797 & 77,493 &	1,059,729 \\
reduction	& 56\% &	51\% &	56\% &	52\% &	51\% &	48\% &	53\% &	52\% &	51\% &	59\% &	- & 51\%  \\
\hline

\multicolumn{13}{c}{in-domain collapsing without coreference resolution}\\
KG concepts &	184,218 &	199,894 &	234,399 &	31,525 &	32,937 &	83,445 &	80,476 &	56,690 &	14,911 &	284,547 & - &	1,203,042 \\
reduction &	45\% &	46\% &	45\% &	31\% &	29\% &	35\% &	37\% &	34\% &	27\% &	51\% &	-& 45\% 

\end{tabular}
}
\vspace{-1em}
\end{table}

\subsubsection{Evaluation of KG Population Strategies:}
Next, we discuss the different KG population strategies.
For each strategy, Table~\ref{tab:kg_concepts_per_domain_type} reports the number of concepts in the populated KG and the percentage reduction of mentions to concepts, and in Table~\ref{tab:results_collapsing_methods} the evaluation results of KGs against the Test-STM-KG.

\paragraph{Cross-Domain vs. In-Domain Collapsing:}
Cross-domain collapsing achieves a higher CoNLL F1 score of 64.8 than in-domain collapsing with a score of 63.5 (see Table~\ref{tab:results_collapsing_methods}).
However, in-domain collapsing yields (as expected) a higher precision (CoNLL P 85.5), since some terms have different meanings across domains (e.g. \textit{Measure\_(mathematics)} vs. \textit{Measurement} in https://en.wikipedia.org).
Furthermore, the Test-STM-KG has only 31 cross-domain concepts due to its small size. 
Thus, we expect that cross-domain collapsing would yield worse results on a larger test set. 

Furthermore, as shown in Table~\ref{tab:kg_concepts_per_domain_type}, cross-domain collapsing yields less concepts than in-domain collapsing (more than 994,000 versus 1.1 Mio. concepts).
We can also observe that only 70,044 (7\%) of the concepts are used across multiple domains.
This indicates, that each scientific domain mostly uses its own terminology.
However, the concepts used across domains can have different meanings. 
Thus, when precision is more important than recall in downstream tasks, in-domain collapsing should be the preferred choice.

\paragraph{Effect of Coreference Resolution:}
Coreference resolution has only a small impact on the number of resulting concepts in a populated KG (see Table~\ref{tab:kg_concepts_per_domain_type}).
However, as shown in Table~\ref{tab:results_collapsing_methods}, leaving out the coreference resolution step during KG population yields only low CoNLL F1 scores, i.e. 41.7 (-21.8) F1 and 43.5 (-21.3) F1. 
Thus, coreference resolution significantly improves the quality of a populated KG .

\begin{table}[tb]
\caption{Performance of the collapsing strategies evaluated against the \emph{Test-STM-KG}: in-domain and cross-domain collapsing  with and without coreference resolution.}
\label{tab:results_collapsing_methods}
\resizebox{\textwidth}{!}{
\begin{tabular}{l|r|rrr|rrr|rrr|rrr}
               & \#concepts & \multicolumn{3}{c}{$MUC$} & \multicolumn{3}{c}{$B^3$} & \multicolumn{3}{c}{$CEAFe_{\phi 4}$} & \multicolumn{3}{c}{$CoNLL$} \\
               & in KG          & P      & R      & F1    & P      & R     & F     & P       & R      & F1     & P       & R      & F1     \\ \hline
in-domain collapsing & 859       & \textbf{86.3}   & 70.6   & 77.7  & \textbf{86.0}   & 69.0  & 76.6  & 84.1    & 23.1   & 36.2   & \textbf{85.5}    & 54.2   & 63.5   \\
- without coreferences &	900	& 75.5 & 38.8 &  51.2 &  75.2 & 37.9 & 50.4 & 71.1 & 14.0 & 23.4 & 73.9 & 30.2 & 41.7   \\ \hline
cross-domain collapsing  & 837       & 85.0   & \textbf{73.0}   & \textbf{78.5}  & 84.5   & \textbf{72.1}  & \textbf{77.8}  & \textbf{84.7}    & \textbf{24.6}   & \textbf{38.1}   & 84.7    & \textbf{56.6}   & \textbf{64.8}  \\
- without coreferences & 876	& 73.5 & 	41.0 & 	52.6 & 	72.2 & 	15.5 & 	25.5 & 	72.2 & 	15.5 & 	25.5 & 	73.0 & 	32.4 & 	43.5  \\ 
\end{tabular}
}
\vspace{-1em}
\end{table}

\subsubsection{Qualitative Analysis:}
We also inspected the top five frequent domain-specific concepts in the populated KG (a list of these concepts can be found in our public repository). As far as we can judge with our computer science background, we consider the extracted top frequent concepts to be reasonable and useful for the domains. For instance, in \textit{Ast}, the method `standard model' is frequently mentioned, while in \textit{CS} the process `cyber attack' appears most often.
The frequency of the top concepts differs significantly between the domains: In \textit{Med}, \textit{Ast}, \textit{Eng}, \textit{ES} and \textit{Agr}, a top frequent concept is referenced 10.8, 10.2, 4.9, 3.8, and 3.1 times per 1000 abstracts, respectively. In \textit{Che}, \textit{MS}, \textit{Mat}, \textit{Bio}, and \textit{CS}, a top frequent concept is referenced only by few abstracts (0.3, 0.4, 1.0, 1.4, and 2.3, respectively, per 1000 abstracts).


\section{Conclusions}
In this paper, we have investigated the task of coreference resolution in research papers across 10 different scientific disciplines. We have annotated a corpus that comprises 110 abstracts with coreferences with a substantial inter-coder agreement. Our baseline results with current state-of-the-art approaches for coreference resolution demonstrate that current approaches perform poorly on our corpus. The proposed approach, which uses sequential transfer learning and exploits annotated datasets from the general domain, outperforms noticeably the state-of-the-art baselines. Thus, our transfer learning approach can help to reduce annotation costs for scientific papers, while obtaining high-quality results at the same time. 

Furthermore, we have investigated the impact of coreference resolution on KG population. For this purpose, we have compiled a gold KG from our annotated corpus and propose an evaluation procedure for KG population strategies. We have demonstrated that coreference resolution has a small impact on the number of resulting concepts in the KG, but improved significantly the quality of the KG.
Finally, we have generated a research KG from 55,485 abstracts of the 10 investigated domains. 
We show that each domain mostly uses its own terminology and that the populated KG contains useful concepts.
To facilitate further research, we make our corpora and source code publicly available: \url{https://github.com/arthurbra/stm-coref}

In future work, we plan to evaluate multi-task learning approaches, and to populate and evaluate a much larger research KG to get more insights in scientific language use.

%
%
%
\bibliographystyle{splncs04}
\bibliography{references}
\end{document}